\documentclass{aa}  

\usepackage{graphicx}
\usepackage{natbib}
\usepackage{txfonts}
\usepackage{textcomp}
\bibpunct{(}{)}{;}{a}{}{,}             
\begin{document} 

   \title{Comparing the properties of the X-shaped bulges of NGC~4710 and the Milky Way with MUSE\thanks{Based on observations collected at the ESO La Silla-Paranal Observatory within MUSE science verification program 60.A-9307(A).}}


   \author{O. A. Gonzalez\inst{1,2}
          \and 
          D. A. Gadotti\inst{1}
          \and
          V. P. Debattista\inst{3}
          \and
          M. Rejkuba\inst{4, 5}          
          \and 
          E. Valenti\inst{4}
          \and      
          M. Zoccali\inst{6,8}
          \and
          L. Coccato\inst{4}
          \and
          D. Minniti\inst{7, 8, 9}
          \and
          M. Ness\inst{10}
           }

   \institute{European Southern Observatory, Ave. Alonso de Cordova 3107, Casilla 19, 19001, Santiago, Chile \\
   	   \email{ogonzale@eso.org}
	\and
	    Institute for Astronomy, University of Edinburgh, Royal Observatory, Blackford Hill, Edinburgh, EH9 3HJ,  UK
    \and
	    Jeremiah Horrocks Institute, University of Central Lancashire, Preston PR1 2HE, UK
	\and
   	    European Southern Observatory, Karl-Schwarzschild Strasse 2, D-85748 Garching, Germany
         \and
           Excellence Cluster Universe, Boltzmannstr. 2, D-85748, Garching, Germany
          \and
           Instituto de Astrof\'{\i}sica, Facultad de F\'{\i}sica, Pontificia Universidad Cat\'olica de Chile, Av. Vicu\~na Mackenna 4860, Santiago 22, Chile
          \and
          Departamento de Ciencias F\'isicas, Universidad Andr\'es Bello, Rep\'ublica 220, Santiago, Chile
          \and
          The Milky Way Millennium Nucleus, Av. Vicu\~na Mackenna 4860, 782-0436, Macul, Santiago, Chile
          \and
           Vatican Observatory, V00120 Vatican City State, Italy.   
           \and
           Max-Planck-Institut fur Astronomie, Konigstuhl 17, D- 69117 Heidelberg, Germany
          }

   \date{Received ; accepted }

 
  \abstract
   {Our view of the Milky Way's structure, and in particular its bulge, is obscured by the intervening stars, dust, and gas in the disc. While great progress in understanding the bulge has been achieved with past and ongoing observations, the comparison of its global chemodynamical properties with respect to those of bulges seen in external galaxies remains to be done.}
{We used the new ESO VLT instrument MUSE to obtain spectral and imaging coverage of NGC~4710. The wide area and excellent sampling of the MUSE integral field spectrograph allows us to investigate the dynamical properties of the X-shaped bulge of NGC~4710 and compare it with the properties of the Milky Way's own X-shaped bulge.}
{We measured the radial velocities, velocity dispersion, and stellar populations using a penalized pixel full spectral fitting technique adopting simple stellar populations models, on a $1' \times 1'$ area centred on the bulge of NGC~4710. We have constructed the velocity maps of the bulge of NGC~4710 and we investigated the presence of vertical metallicity gradients. These properties were compared to those of the Milky Way bulge and as well as to a simulated galaxy with boxy/peanut bulge.
   }
{We find the line-of-sight velocity maps and 1D rotation curves of the bulge of NGC~4710 to be remarkably similar to those of the Milky Way bulge. Some specific differences that were identified are in good agreement with the expectations from variations in the bar orientation angle. The bulge of NGC 4710 has a boxy-peanut morphology with a pronounced X-shape, showing no indication of any additional spheroidally distributed bulge population, in which we measure a vertical metallicity gradient of 0.35~dex/kpc.}
 {The general properties of NGC 4710 are very similar to those observed in the MW bulge. However, it has been suggested that the MW bulge has an additional component that is comprised of the oldest, most metal-poor stars and which is not part of the boxy-peanut bulge structure. Such a population is not observed in NGC 4710, but could be hidden in the integrated light we observed.}

\titlerunning{Kinematical comparison of the X shape bulge of NGC~4710 and the Milky Way}
\maketitle
%

\section{Introduction}

The understanding of galactic bulges is a basic step to unveiling the formation history and evolution of galaxies. By mapping the kinematics, stellar population, and the morphological signatures of its bulge, it is possible to constrain the history of events that occurred during the assembly of the galaxy.

Bulges are generally classified into three distinct groups: I) classical bulges, which are spheroidal components dominated by velocity dispersion of old, alpha-enhanced stars as expected from a fast and early formation via dissipative collapse or mergers \citep[e.g.][and references therein]{brook-christensen+15}; II) pseudo-bulges, which are disc-like, rotation dominated structures populated by young, metal-rich stars, formed from the inflow of gas to the center of the galaxy due to the influence of a bar \citep[][]{kormendy+13}; and III) Boxy/Peanut (B/P) bulges which are the result of buckling bar instabilities that favour the heating of the stellar orbits along the vertical direction resulting in a thick structure that swells up from the disc in a peanut or X-shape\footnote{We note that whether these structures appear boxy, peanut or X shaped, depends on projection effects and the strength of the buckling instability. We therefore indistinguishably use the terms B/P and X-shape bulge here to refer to same structure, i.e. the vertically thickened inner parts of the stellar bar.} \citep[e.g.][]{ComSan81, Ath05b}. B/P bulges are common in disk galaxies, being found in nearly half of edge-on disc galaxies (with a fraction missed due to unfavourable orientation \citep{lutticke+00}) and even our own Galaxy has been recently proven to host one \citep[][and references therein]{wegg-gerhard+13}. However, the stellar population properties of B/P bulges have not yet been entirely defined, mostly due to the difficulties of disentangling the different bulge components in the analysed samples. As recently pointed out by \citet[][]{laurikainen+15}, dedicated studies of individual galaxies, where both the structural decomposition and stellar population analysis can be performed, is fundamental to characterise the properties of B/P bulges.

In the Milky Way, the presence of a bar in the inner regions is now well established \citep{stanek+94} and its boxy shape has been mapped by different IR surveys such as COBE, 2MASS, and most recently the VVV survey \citep{minniti+10}. The radial velocity measurements from spectroscopic surveys, based on hundreds of velocities of M-giants in BRAVA \citep{howard+09} and K-giants in GIBS \citep{zoccali+14}, have provided strong evidence for cylindrical rotation,  the expected kinematical signature of rotating bars \citep[but see][]{WilZamBur11}.  Furthermore, following the boxy shape of the Galactic bar mapped by COBE and 2MASS, the analysis of the distribution of red clump stars from different datasets has provided independent evidence for the X-shaped morphology of the Milky Way \citep{nataf+10,mcwilliam-zoccali+10,saito+11}. Recently, \citet[][]{wegg-gerhard+13} used photometric data from the VVV survey to map the X-shaped bulge in detail and thus reaffirming the B/P nature of the Galactic bulge. Recent stellar population studies have revealed a dominating old (10 Gyr) $\alpha$-enhanced population, as well as the presence of a radial metallicity gradient \citep[e.g.][]{zoccali+03,gonzalez+13,valenti+13}. The presence of young (<5 Gyr), metal-rich stars, particularly at low latitudes, has also been discussed \citep{bensby+13, ness+14, dekany15}. It has been suggested that the observed spatial distribution of [Fe/H], [$\alpha$/Fe], and possibly also stellar ages in the Galactic bulge, could be the result of a contribution of different components dominating at different distances from the Galactic plane, namely the B/P bulge and a classical bulge \citep{babusiaux+10, hill+11, dekany+13}.   

Large new datasets are therefore providing new observational insights into the bulge of our Galaxy. However, a decomposition of the different components of the Milky Way bulge is highly complicated by our location in the disc of the Galaxy as well as a combination of effects due to extinction \citep{gonzalez+12}, line-of-sight depth and bar orientation \citep{wegg-gerhard+13}, and the overall geometry and kinematics of different overlapping Galactic components. It is then left to a reconstruction method, based on the interpretation of the different sets of observations described above. Image decomposition, on the other hand, can be used to study the different bulge components of other galaxies \citep[e.g.][]{gadotti+12, LauSalBut05, LauSalBut10, MenDebCor14}. It is for this reason that it becomes fundamental to provide a link between the techniques used in the study of the unresolved stellar properties of external galaxies and those used in the Milky Way \citep[see][for a recent review on this matter]{gonzalez-gadotti+15}. In this context, NGC~4710 appears as an ideal candidate to test the interpretations of the Milky Way bulge observations.

According to NED and HyperLeda databases, NGC~4710 is an edge-on early-type (S0 or SA(r)0) galaxy in the nearby universe.  It is located in the Virgo cluster, and is included in the ATLAS$^{\mathrm{3D}}$ sample of 260 early-type (E and S0) galaxies. NGC 4710 is unusually gas rich, having the second highest average CO surface density \citep{young+11} and is one of the two galaxies most significantly affected by dust extinction in the sample of \citet{scott+13}.  \citet{krajnovic+13} list it among the sample of 13 "uncertain" galaxies in the early-type ATLAS$^{\mathrm{3D}}$ sample given the strong dust features in the nucleus and uncertain nuclear profile fits. Its ionized, molecular and stellar kinematics  are aligned \citep{davis+11}, and its stellar $V_{rms}=\sqrt{V^2+\sigma^2}$ shows a butterfly-like shape characteristic of galaxies with small bulges \citep{cappellari+13}. 

NGC~4710 can be found among the galaxies sampled by the \textit{Spitzer} Survey of Stellar Structure in Galaxies (S$^4$G) and the recent morphological analysis of \citet{buta+15} classified it as an exactly edge-on barred galaxy with a X-shaped bulge and visible ansae on each side of the center. Furthermore, no co-existing, large-scale classical bulge component has been found based on its light-decomposition analysis \citep{gadotti+12}.

Based on the fact that both the Milky Way and NGC~4710 have a B/P bulge, we have the possibility to build a link between the studies of resolved and unresolved properties of stellar populations of this type of bulges, thus bringing to a common ground our detailed knowledge of the Milky Way bulge and that of unresolved external galaxies. With this aim, in this article we investigate the kinematics and stellar populations of the B/P bulge of NGC~4710 and we compare them directly to those of the Milky Way bulge. This comparison becomes particularly important when considering recent studies which suggest that the spatial distribution of the oldest population of the Milky Way bulge, traced by RR Lyrae stars, follows a spheroidal component, possibly co-existing with the B/P bulge of the Milky Way \citep[][]{dekany+13} \citep[but see also][]{pietr+15}. Therefore, we can also explore the presence of any specific signature, that is not found in the pure B/P bulge of NGC~4710, which can therefore be identified as an additional component in the bulge of the Milky Way.


\section{Observations and data reduction}

The observations of NGC~4710 were taken as part of the MUSE Science Verification observing run in June 2014. MUSE \citep{bacon+10} is an optical wide-field integral field spectrograph installed in UT4 at the ESO Very Large Telescope. It uses the image slicing technique to cover a field-of-view (FOV) of $1\arcmin\times1\arcmin$ in wide-field mode (WFM) resulting in a sampling of $0.2\arcsec\times0.2\arcsec$ spaxels (equivalent to 0.016 \mbox{kpc} at a distance of 16.9 \mbox{Mpc} for NGC~4710). The full field is split up into 24 sub-fields (each $2.5\arcsec\times60\arcsec$ in WFM) which are fed into one of the 24 integral field units (IFUs) of the instrument. In addition, MUSE covers an impressive wavelength range from $4650~\AA$ to $9300~\AA$ at a spectral resolution of R$\sim$2,000 at $4600~\AA$ and R$\sim$4000 at $9300~\AA$.

We observed the bulge of NGC~4710 using MUSE without adaptive optics, at the nominal wavelength range. The central coordinates of the observed field ($\alpha = 12h49m37.9s$, $\delta=+15^{\circ}10'00.8''$, J2000) had an offset with respect to the centre of the galaxy in order to avoid contamination from a bright star in the field-of-view. This configuration allowed us to cover the inner $15\arcsec$ of the X-shape bulge as well as to extend the coverage to the expected limit of the bulge at the NW side of the galaxy. 

Stellar kinematic maps of NGC~4710 have been previously constructed by the SAURON and ATLAS$^{\mathrm{3D}}$ surveys. Although these surveys have provided a large set of homogeneously analysed galaxies, the larger FOV and spatial resolution of MUSE allows for a superior, detailed analysis of the kinematical signatures present in a single galaxy. In this case, the MUSE dataset is superior both with respect to the spatial coverage that allows us to trace the B/P bulge to the external regions of NGC~4710 and a spatial resolution of 0.2\arcsec in the inner regions. 


\begin{figure}
\centering
\includegraphics[width=9cm,angle=0]{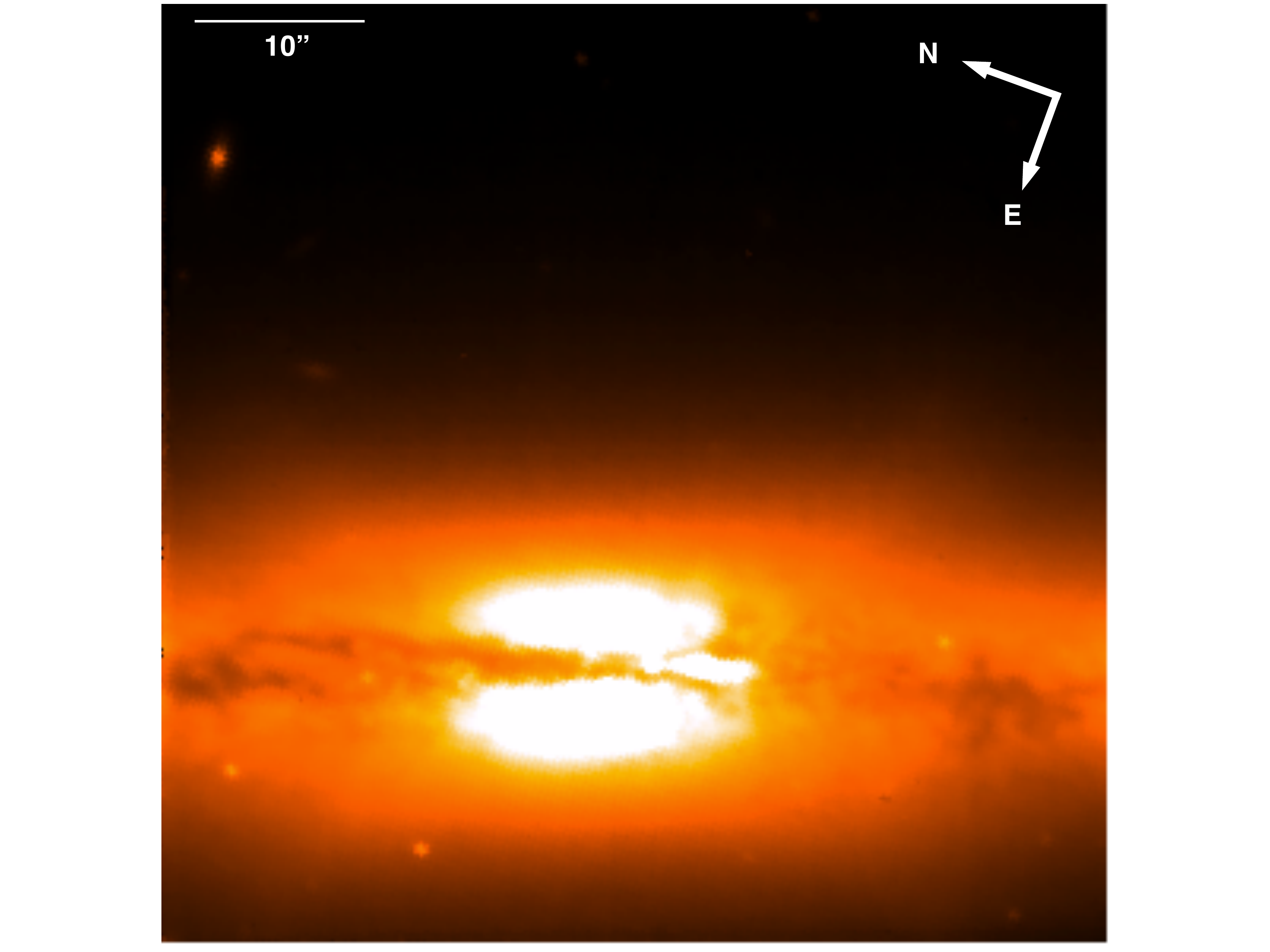}

\caption{Reconstructed image of the MUSE cube resulting from the co-addition of 4 individual exposures mapping the bulge of NGC~4710. The bar in the top left panel indicates a scale of 10\arcsec.}
\label{fov}
\end{figure}

A total of 4 exposures of 600s were obtained, each of which was followed by a sky exposure obtained in an offset field ($\alpha = 12h49m48.5s$, $\delta=+15^{\circ}06'32.1''$, J2000). We adopted an exposure time of 180s for the sky exposures. We iterated between two rotator angle positions of $-28.8^{\circ}$ and $90^{\circ}-28.8^{\circ}=61.2^{\circ}$ between each integration. Figure~\ref{fov} shows the reconstructed image of the MUSE cube; the B/P bulge of NGC~4710 is clearly visible.

The data were reduced using the MUSE pipeline (v1.0) and using the legacy static calibrations provided by ESO for science observed during science verification runs. The reduction process of the individual scientific exposures was performed executing the specifically designed MUSE pipeline recipes. The final data-cube analysed here corresponds to the combined outcome of the four individual observations. Sky subtraction was performed using our dedicated sky observations using the subtract-model method and a sky fraction of 85$\%$ to account for possible contamination from the foreground and faint halo of NGC~4710 which could still be present in our offset field. Astrometric calibrations were calculated by the pipeline using the static calibration database. Telluric correction and flux normalisation was performed using observations of standard star GD153 observed by the MUSE-SV team immediately before the start of our observations.  For the present study, we have only focused on the spectral region between $4750$ -- $6100~\AA$ and for this reason the pipeline reduction procedure was restricted to this region. This limited wavelength coverage facilitated the complicated sky subtraction and telluric correction arising when reducing the complete MUSE wavelength region. 

\begin{figure}[]
\centering
\resizebox{\hsize}{!}{
\includegraphics[scale=1]{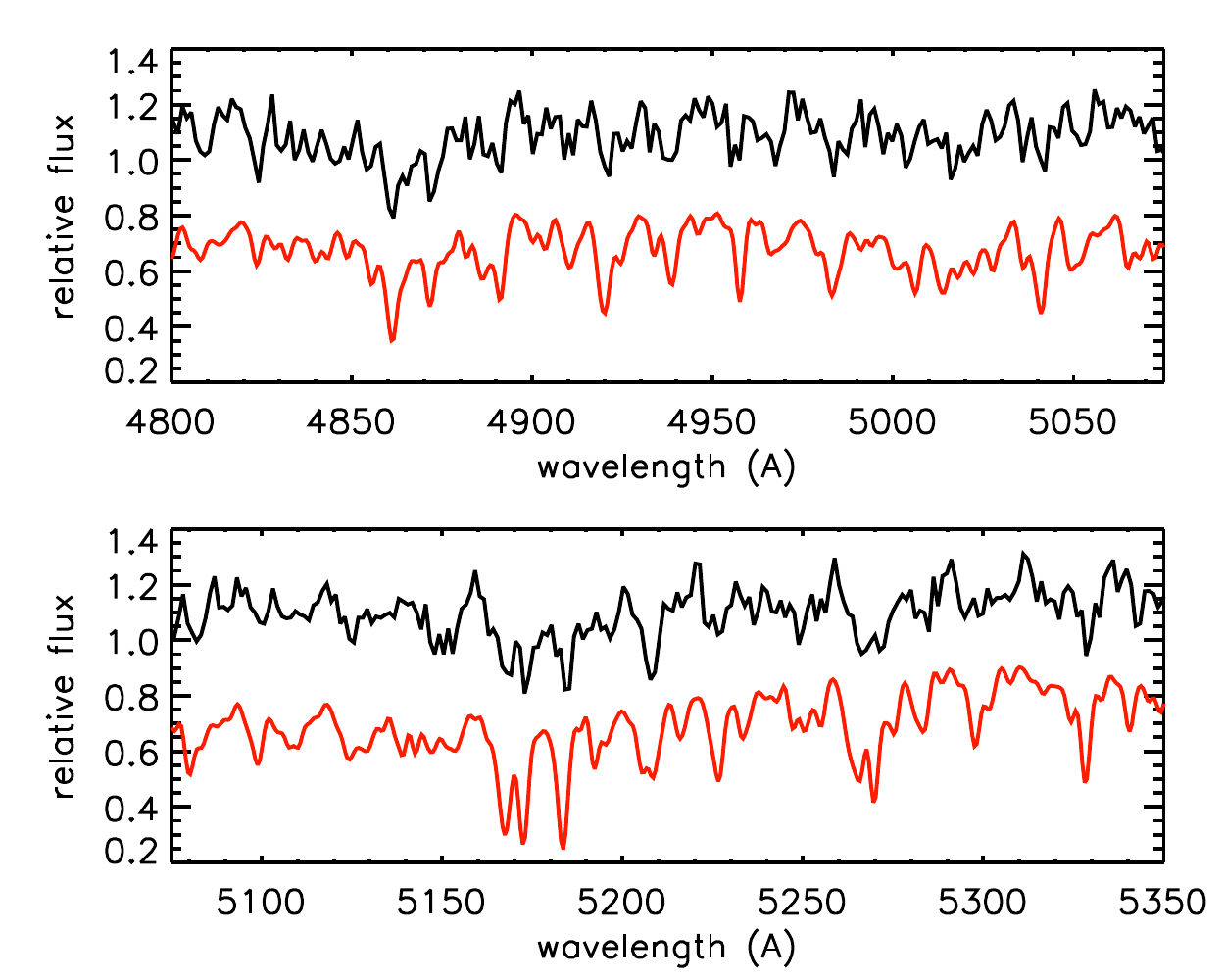} 
}
\caption{Example spectrum of NGC~4710 corresponding to one bin resulting from the Voronoi binning with signal-to-noise 50 (black line). The displayed spectral region corresponds to the one used on the penalized pixel fitting. A template spectrum from the \citet{vazdekis+10} library is also included as a reference (red line).}
\label{spectra}
\end{figure}

Once the final cube was constructed, the next step was to use the Voronoi binning method of \citet{capellari-copin+03} to spatially bin the cube in order to maximise the spatial resolution while providing a minimum signal-to-noise (S/N) ratio of 50 in each spatial bin. Setting this S/N limit allows for a proper analysis of the stellar kinematics of the bulge of NGC~4710 to be carried with confidence in our entire field-of-view. Figure~\ref{spectra} shows an example of the spectra corresponding to one spatial bin of the resulting MUSE cube.

The kinematic analysis of the resulting spectra in each spatial bin was carried using the penalized pixel fitting (pPXF) routines in IDL developed by \citet{capellari-emsellem+04} using the Single Stellar Population library of \citet{vazdekis+10} as reference. This template library has the same resolution as the analysed spectral range covered by MUSE (FWHM$\sim$2.3\AA) and thus the convolution of the template spectra was not necessary. Gas emission lines are masked-out from the fit by pPXF which then evaluates the galaxy stellar kinematics by fitting the templates to the observed spectrum in pixel space using a maximum penalized likelihood method. The first four terms of the Gauss-Hermite series are then extracted from fitting the line-of-sight velocity distribution (LOSVD), providing the LOS mean velocity V, the velocity dispersion $\sigma$, and the next two Gauss-Hermite coefficients h3 and h4.  The LOS mean velocity and velocity dispersion ($\sigma$) are used to construct rotation maps that can be directly compared to those constructed in the Milky Way bulge based on the GIBS survey data \citep{zoccali+14}. We do not include the h3 and h4 moments in our analysis because MUSE has an instrumental Line Spread Function (LSF) which is strongly not-Gaussian. Moreover, the LSF varies on a spaxel-by-spaxel basis and also has some wavelength dependence. Therefore, the h3 and h4 measurements do not contain only the information on the galaxy's orbital distribution, but are also contaminated by the MUSE instrumental LSF. This contamination is negligible in the "hot" regions, dominated by velocity dispersion (e.g. $\sigma$ > 90 km/s), but become more important in those regions characterized by $\sigma$ similar to the instrumental $\sigma$ of MUSE (e.g outer regions, where $\sigma \sim$50 km/s). It is however important to keep h3 and h4 in the fit performed by pPXF, because otherwise all the instrumental effects will contaminate our measurements of V and $\sigma$. The typical errors in our V and $\sigma$ measurements are 6 and 8 km/s, respectively.

\begin{figure*}[]
\centering
\includegraphics[width=15.5cm,angle=0,trim=0.0cm 10.5cm 0cm 0cm, clip=true]{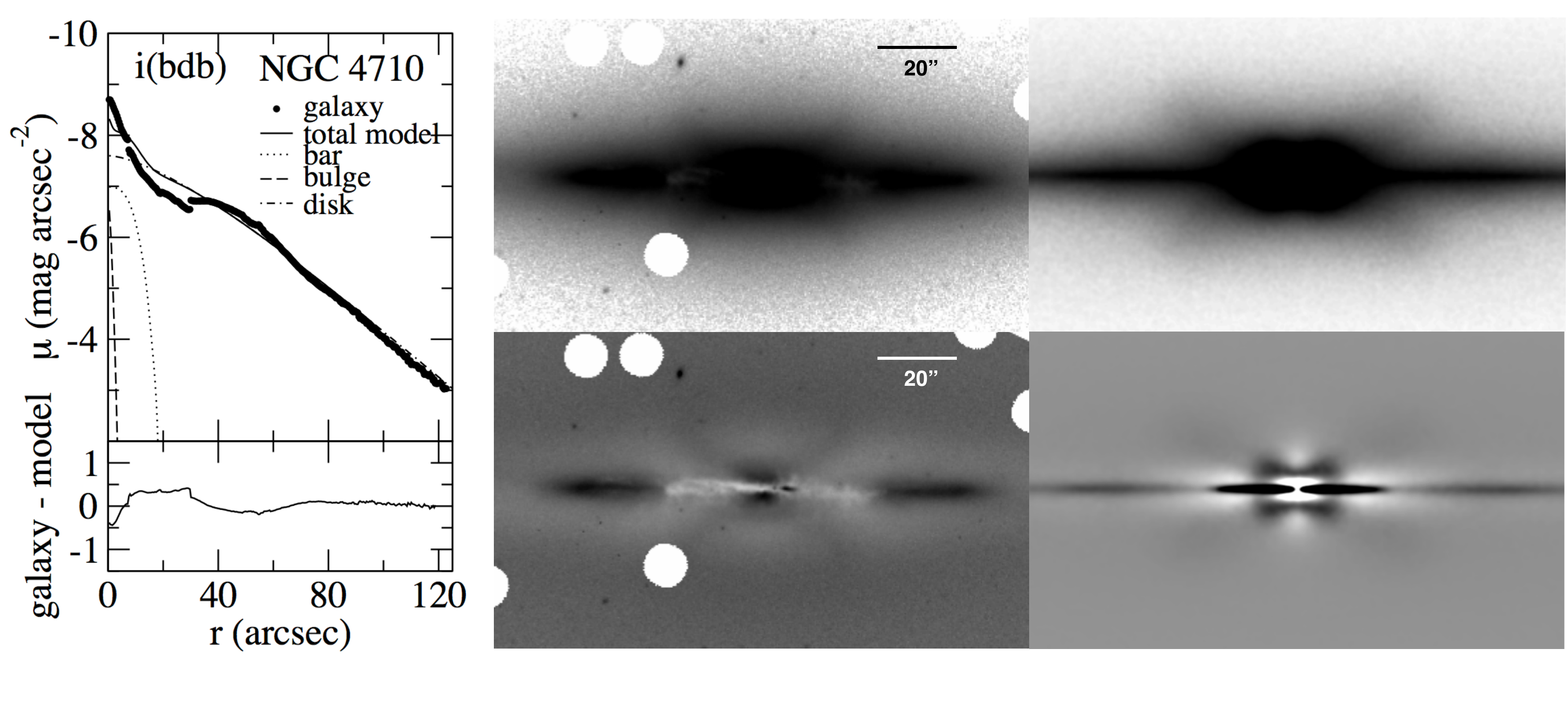}\\
\caption{The left panel shows the light profile for NGC~4710 as well as the contribution of the best fitting components from the BUDDA model light decomposition. The central upper panel shows the SDSS i-band image of NGC~4710 used for the structural decomposition. The corresponding residual image of the decomposition for NGC~4710 is shown in the central lower panel. The right panels show the snapshot image of the galaxy simulation (upper panel) and the residual image from its structural decomposition (lower panel).}
\label{budda_4710}
\end{figure*}

\section{Structural components in NGC~4710}

In order to correctly interpret the rotation and velocity dispersion maps, it is crucial to understand the structural composition of the sampled region of NGC~4710. In this way we know which component (or components) dominates the spectra in each region. We thus refer to the analysis of \citet{gadotti+12} where a decomposition of NGC~4710 was performed using BUDDA \citep{deSGaddos04, gadotti+08} to decompose the $i$-band image of NGC~4710 obtained from the Sloan Digital Sky Survey (SDSS) into different structural components.

Figure~\ref{budda_4710} shows the best fitting light profile for NGC~4710 from \citet{gadotti+12}. The best model for NGC~4710 includes a vertically extended bar structure. On the other hand, the best fitting model includes a small structure, with a S\'ersic index of 0.7, that contributes only with a very minor fraction, 0.1$\%$, to the total luminosity of the galaxy. Its near-exponential light profile suggests an origin connected to the main exponential disc, so it is likely a structural component built from disc material. Figure~\ref{budda_4710} shows the residual image after the BUDDA model subtraction. The clear X-shape obtained in the residual image strongly corroborates that the bulge of NGC~4710 belongs to the family of B/P bulges.

In addition, the residual image in Fig.~\ref{budda_4710} also shows clearly a few substructures that may be associated to the bar. Firstly, as seen in \citet{gadotti+12}, beyond the dust lane, at each side of the centre, a narrow excess of light indicates the presence of a density enhancement in the disk that can be the bar ansae, spiral arms or a ring. These structures appear to be common in edge-on galaxies with B/P bulges \citep[e.g.][]{BurAroAth06}, and in simulations of barred galaxies when viewed at an edge-on projection \citep[e.g.][]{laurikainen+15}.

\begin{figure*}[]
\centering
\includegraphics[width=18.5cm,angle=0,trim=1.3cm 0.5cm 0cm 0cm, clip=true]{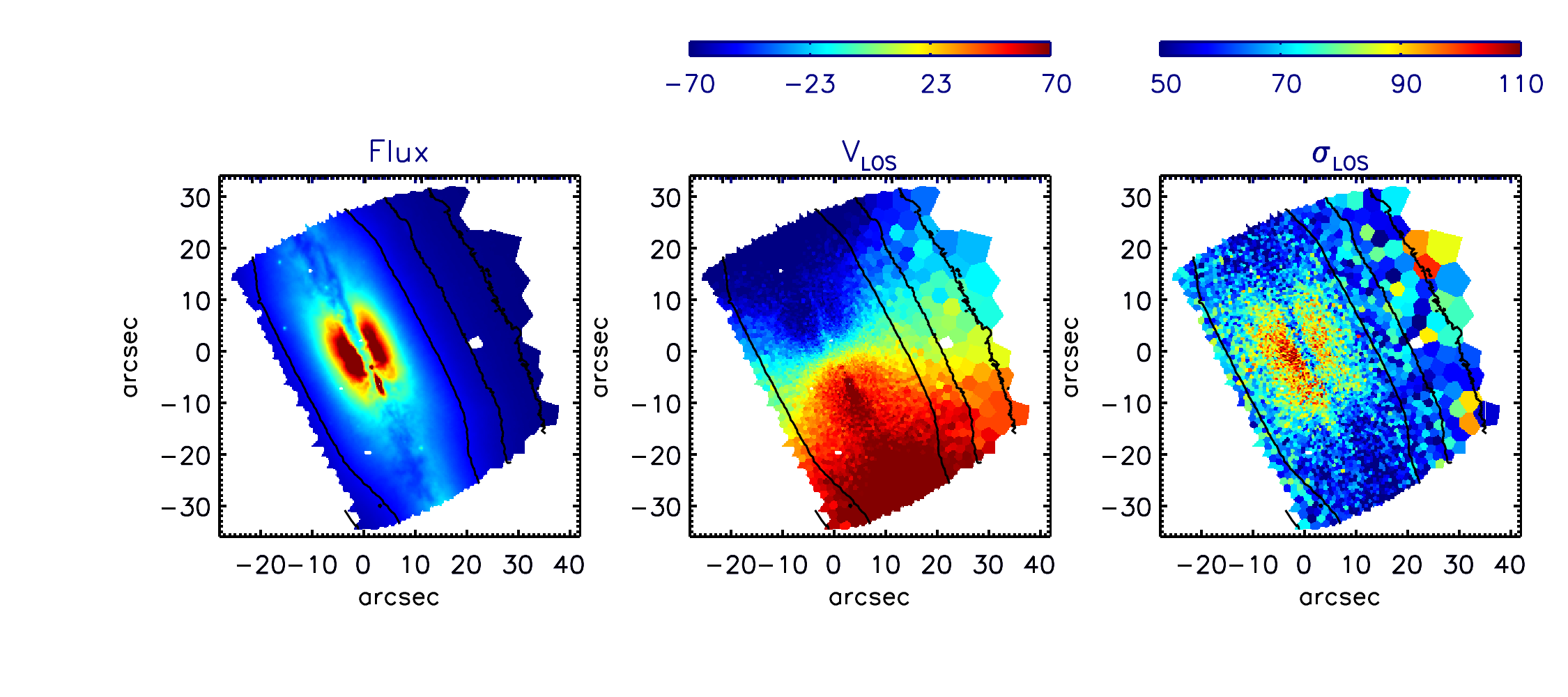}\\
\caption{The intensity map of our NGC~4710 MUSE field is shown in the left panel. LOS velocity and velocity dispersion maps of NGC~4710 are shown in the upper middle and right panels, respectively. Each map has the corresponding isophotes over-plotted, spaced by one magnitude.}
\label{NGC4710kine}
\end{figure*}

Secondly, in the central $\sim5\arcsec$ of the residual image one notices a small structure that sticks out prominently from the disc plane (although still significantly less extended that the B/P bulge in the vertical direction). This structure has a pronounced boxy shape, which hints at it being associated to the large-scale B/P. Although the model fitted has a component to account for the B/P, it is designed to account for its average light distribution, so deviations from the average will stand out in the residual image. That would be the case if the central light distribution exceeds what is expected in the global B/P model, producing the observed residual image. In order to test whether this central structure is just part of the B/P, similarly to the X-shape residual component, we performed a structural analysis using BUDDA on a snapshot of the galaxy simulation of \citet{cole+14}, as was done with the SDSS image of NGC~4710. The simulation contains only a disc and bar (plus B/P) components, and the snapshot was convolved with a circular Gaussian function to mimic PSF effects. The Gaussian FWHM was chosen to reproduce the same relative PSF as in the SDSS image of NGC~4710. As found for NGC~4710, the best BUDDA model essentially contains only a disc and a bar component. The residual image derived from this fit is shown on the bottom right of Fig.~\ref{budda_4710}. One clearly sees both the large-scale X-shape residual component and a central smaller structure very similar to that found in the residual image of NGC~4710. However, in this case, the central structure is most likely associated to the B/P, as indicated by the conspicuous X-shape it has. We conjecture that the corresponding X-shape of the central structure is not entirely discernible in the residual image of NGC~4710 due to a combination of projection effects and spatial resolution of the SDSS image. If the galaxy is not seen at a perfect side-on projection the smaller, innermost X-shape structure may appear as boxy only.

Thus, in the case of NGC~4710, the structural analysis indicates that our kinematic study will be tracing the dynamical signatures dominated by the disk and bar, in the regions near the plane of the galaxy, and that of the associated B/P bulge at increasing height.
\section{Kinematic maps of NGC~4710}

We have used the resulting kinematics derived from the absorption-line fitting with pPXF to produce kinematic maps of NGC~4710. Figure~\ref{NGC4710kine} shows the intensity map obtained from collapsing our data-cube across the spectral dimension, as well as the maps of LOS velocity and velocity dispersion that are produced after correcting for the corresponding redshift of NGC~4710 (V=1174 km/s). The isophotes obtained from the reconstructed image are over-plotted in each map and are spaced by one magnitude intervals. The kinematic PA of the galaxy was found to be $24.8\pm3.1$ deg., obtained using the method described in Appendix C of \citet{krajnovic+06} and implemented in the IDL routine FIT\_KINEMATIC\_PA by M. Capellari.

The LOS velocity map shows the signatures of cylindrical rotation, as expected for a B/P bulge, consistent with the photometric analysis of NGC~4710. In theory, a bulge rotating perfectly cylindrically would have a mean LOS velocity, at a given distance from the centre, which is independent of its height above the plane \citep{Saha+13}. A very useful criteria to quantify the deviations from pure cylindrical rotation was provided by equation 2 of \citet{Saha+13}:

\begin{equation}
\delta_{CL}(z,X_j) = \frac{\int_0^{X_j}{\Sigma_{los}(x,z) V_{los}(x,z_i) x^2
dx}}{V_{los}(R_{{b},{1/2}},z\simeq 0) {\int_0^{X_j}{\Sigma_{los}(x,z) x^2 dx}}}.
\label{eq:delCL}
\end{equation}

\noindent where $\Sigma_{los}(x,z)$ is the surface density, $V_{los}(x,z)$ is the LOS velocity, $X_j$ is the projected distance from the minor axis for a slit at a height $z = z_i$, and $V_{los}(R_{{b},{1/2}},z\simeq 0)$ is the velocity at the bulge half-mass radius close to the disc mid-plane. We then fit a straight line to
the ({$\delta_{CL}$ , $z$}) curve and derive the slope of this relation $m_{CL}$. \citet{Saha+13} then defines the degree of cylindrical rotation in the B/P bulge as:

\begin{equation}
\delta_{CL}^n (X_j) = 1 + m_{CL}.
\end{equation}  

\noindent so that a value of $\delta^{n}_{CL}=1$ corresponds to perfect cylindrical rotation while values of $\delta^{n}_{CL}<0.75$ correspond to non-cylindrical rotation.

We applied this equation to our NGC~4710 dataset measuring the surface density and LOS velocity across fixed heights from the plane. Within the half-mass bulge radius of NGC~4710, which corresponds to $\rm R_{b,1/2}$ = $1.35R_e$ = 10\arcsec \citep[where R$_e$=7.4\arcsec,][]{gadotti+12}, we obtain a value of $\delta^{n}_{CL}=0.9$. The slight deviation from cylindrical rotation seen in NGC~4710 could be a consequence of either the presence of a rotating classical bulge or the effect of the bar --and associated B/P -- orientation angle. Following the structural decomposition of NGC~4710 of \citet{gadotti+12}, the presence of a classical bulge in this galaxy can be safely discarded and thus the deviation from cylindrical rotation is likely to be the result of the position angle of the bar. The impact of the bar viewing angle in the measured rotation of the associated B/P bulge has been discussed in \citet{combes+90} and \citet{Athanassoula+02}, and recently investigated in detail by \citet{iannuzzi15}, and is consistent with the very small deviation from cylindrical rotation we see in the B/P bulge of NGC~4710. 

The general behaviour of the maps is in good agreement with the expectations from N-body models presented in \citet{iannuzzi15} for edge-on galaxies hosting B/P bulges. The velocity dispersion increases noticeably towards the inner regions of the B/P bulge. We note that the velocity dispersion map shows an asymmetric profile with a higher velocity dispersion towards the North-East side of the galaxy. This most likely is the consequence of the disk of NGC~4710 not being perfectly edge-on. 

Additionally, we note a very bright spot-like region located in the disc plane of NGC~4710, clearly seen in Fig.~\ref{NGC4710kine} at (X,Y)=($4\arcsec$, $-6\arcsec$). This brighter region could be the result of a dust-free patch in the disc that might allow us to see deeper through the disc up to the inner regions of the bulge. However, the radial velocity map shows a peak in mean velocity for this region which can be interpreted as a signature of this being a region located in the near side of the disc. Furthermore, the spectra show an increase of emission lines in this region, suggesting that this is a gas-rich region in the disc.


\section{Stellar populations in the bulge of NGC~4710}

The stellar populations of bulges and their connection with the processes involved in the bulge formation remains a matter of debate. The different types of bulges lead to expectations on their stellar population content based on their formation scenario. However, the observations of bulges, including that of the Milky Way, show a variety of stellar ages and metallicities that do not always match the expectations of their classification based on their morphology and/or kinematics. Different models and bulge formation theories have explained this lack of a one-to-one relation between the morphological classification of bulges and their observed stellar population properties based on the possibility of a given bulge having a mixture of components, each with its own formation history. Similarly, a variety of properties can be expected for any given component, depending on the characteristics of the processes involved. For example, a wide range of stellar ages can be found even among B/P bulges. Stellar bars in massive galaxies, such as the one in NGC 4710 and in the Milky Way, are expected to already have formed at redshift $~\sim$1 \citep[][]{SheMelElm12} or earlier \citep[][]{simmons+14, gadotti+15}. The buckling instability of bars, and thus the formation of the B/P bulge, can take place on very short time-scales \citep[$\sim$1 Gyr, ][]{Ath08}, opening the possibility of having mostly old stellar ages in B/P structures. On the other hand, a fraction of younger stars can still be present in B/P bulges, provided that star formation has continued in the disc, within the radius at which the bar ends so that these younger stars can be captured by the bar potential. As a consequence, this young population of stars would be expected to be predominantly near the plane \citep{ness+14, dekany15}. 

In this context, NGC 4710, which hosts only a pure B/P bulge, is an excellent laboratory to investigate directly the stellar population properties that can be found in such structures. With this aim we used the Full-Spectrum Fitting capabilities of pPXF\footnote{See http://www-astro.physics.ox.ac.uk/~mxc/software/ for a detailed description of pPXF and the implementation of regularisation when assigning weights to the library templates when performing the penalized fitting procedure.} \citep{capellari-emsellem+04} to investigate the distribution of the mass fraction in the bulge of NGC 4710 in terms of age and metallicity.  A grid of 156 MILES model spectra from \citet{vazdekis+10} (26 ages, 6 [M/H]) was used in pPXF to obtain a best-fit spectrum for five representative regions of the bulge. pPXF searches for an optimal solution that is based on the weights applied to each template spectrum in the age-metallicity grid using a regularisation parameter. The regularisation constrains the solution in such a way that the weights assigned to neighbouring age and metallicity templates changes smoothly while being consistent with the observed spectrum. The final representative stellar population of the observed integrated spectrum is then obtained by calculating a weighted average of the metallicities and ages of the grid using the weights calculated by pPXF. We found that these average values have a negligible sensitivity to the selected regularisation parameter. Thus, we used the default regularisation parameter of 250, which corresponds to a parametrisation error consistent with the typical weights assigned to the templates.

\begin{figure}[]
\centering
\includegraphics[width=9cm,angle=0]{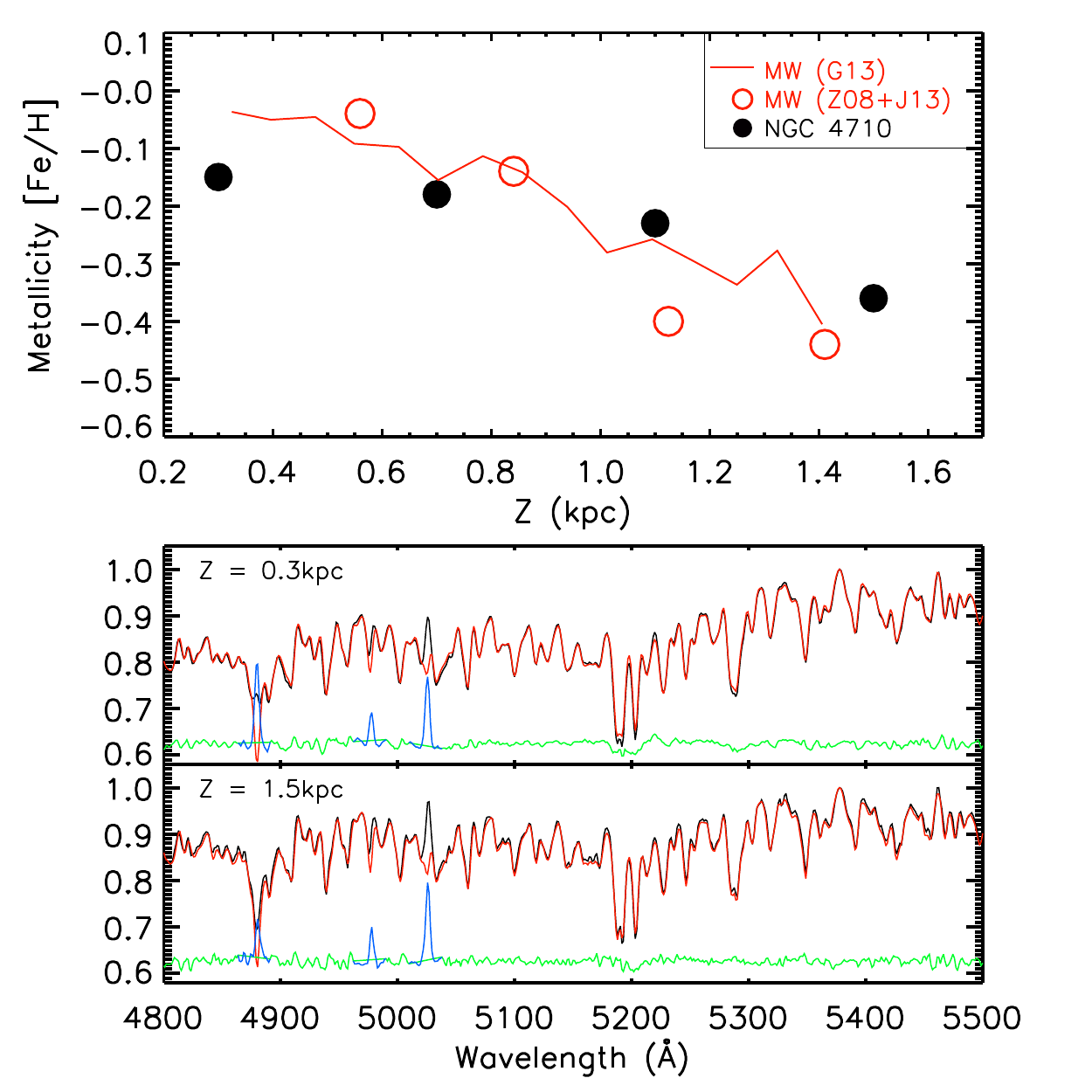}
\caption{pPXF stellar population solution from the full spectral fitting procedure in four regions along the minor axis of NGC~4710. The upper panel shows the mean metallicity of the bulge of NGC~4710 at different distances from the plane (Z) as black filled circles. Mean metallicity measurements for the bulge of the Milky Way are overplotted. The photometric measurements from Gonzalez et al. (2013) are shown as a red solid line, while the spectroscopic measurements from Zoccali et al. (2008) and Johnson et al. (2013) are shown as empty red circles. The two lower panels show the observed (black solid line) and best fitting spectrum (red solid line) of the inner and outermost regions of the bulge of NGC~4710 as well as the masked emission line regions (blue solid lines) and the fitting residuals (green).}
\label{pops}
\end{figure}

We applied the spectral fitting procedure to the integrated spectra of four fields located along the minor axis of the B/P bulge of NGC~4710, specifically at heights from the plane of 0.3, 0.7, 1.1, and 1.5 kpc and derived its dominant age and metallicity. All four fields appear to be consistently dominated by a relatively old population ($>5$ Gyr) with a mean age of $\sim$9 Gyr. On the other hand, metallicity decreases as a function of height from the plane of the galaxy. Figure~\ref{pops} shows the metallicity of the bulge of NGC~4710 in each field compared to the value of the Milky Way bulge at an equivalent height from the plane. As a reference for the minor axis of the Milky Way bulge we use the photometric metallicity maps of \citet{gonzalez+13} and the spectroscopic measurements from \citet{zoccali+08} and \citet{johnson+13}. 

The stellar population analysis of the B/P bulge of NGC 4710 shown in Fig.~\ref{pops} suggests a scenario where NGC~4710 has relatively old stellar ages and a vertical metallicity gradient comparable to that in the bulge of the Milky Way. The metallicity gradient in the bulge of NGC~4710 is an important observational confirmation of vertical gradients reproduced in the simulations of pure B/P bulges seen edge-on \citep[][]{bekki+11, martinez-valpuesta+13}. The exact value of the gradient must be taken with caution as the innermost field is probably affected by a significant contribution of disk stars. However, we note that the scale height of the disc component in the BUDDA model light decomposition of NGC~4710 is $\sim$0.8 kpc (10.3'' at 16.9 Mpc). Therefore, the disc contamination is expected to be negligible at the heights of the two fields at 1.1 and 1.6 kpc from the plane where a variation in metallicity of 0.14 dex is observed. Detailed maps of [Fe/H], $\alpha$-element, and stellar ages based on the measurement of line-strength indices, will be the subject of a dedicated study (Gonzalez et al., in preparation).

\section{Comparison of the bulge of NGC~4710 and of the Milky Way}

\begin{figure*}[]
\centering
{
\includegraphics[scale=0.53,angle=0,trim=0.5cm 0.5cm 0.5cm 0cm, clip=true]{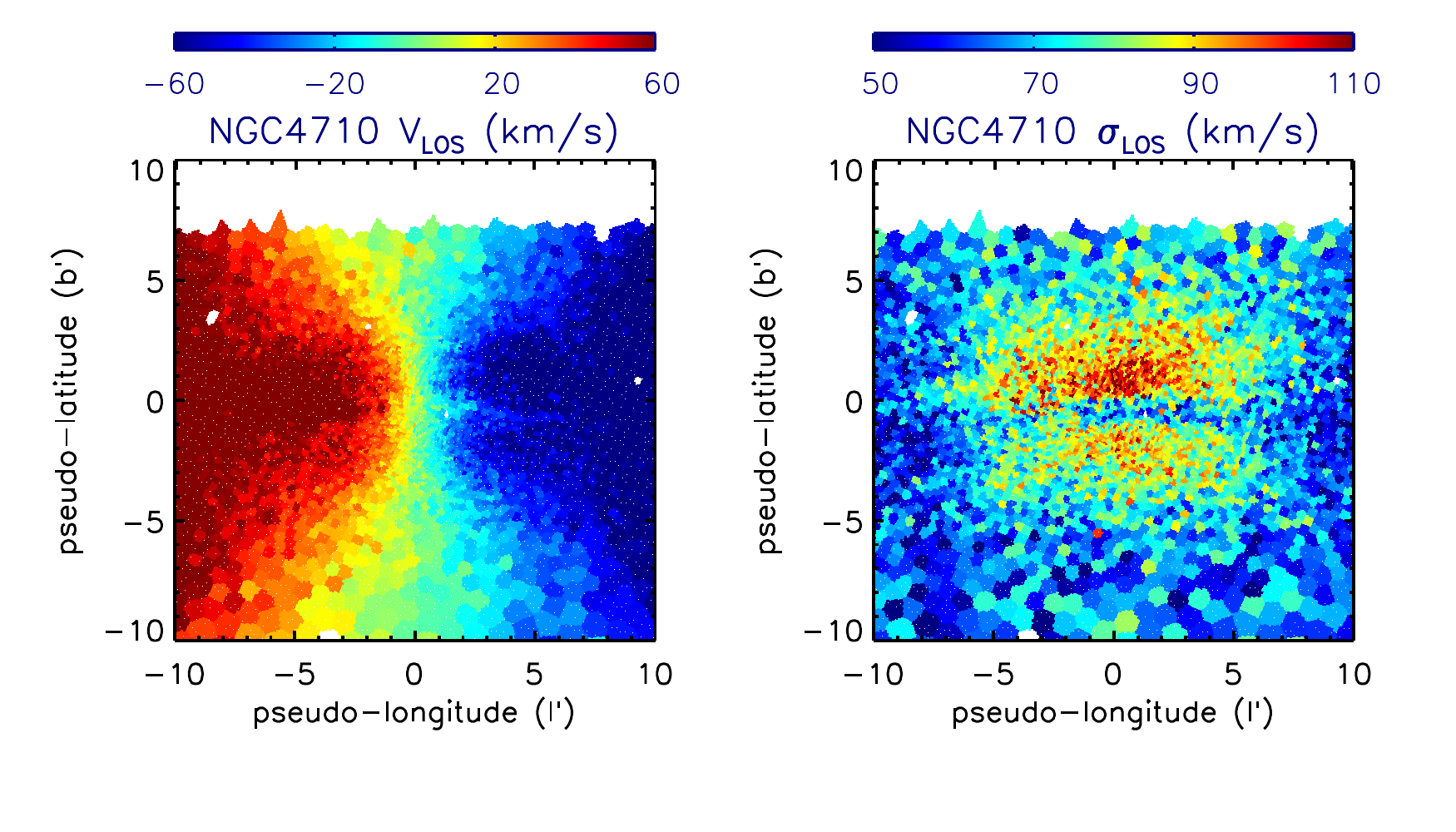}\includegraphics[scale=0.53,angle=0,trim=0.5cm 0.5cm 0.5cm 0cm, clip=true]{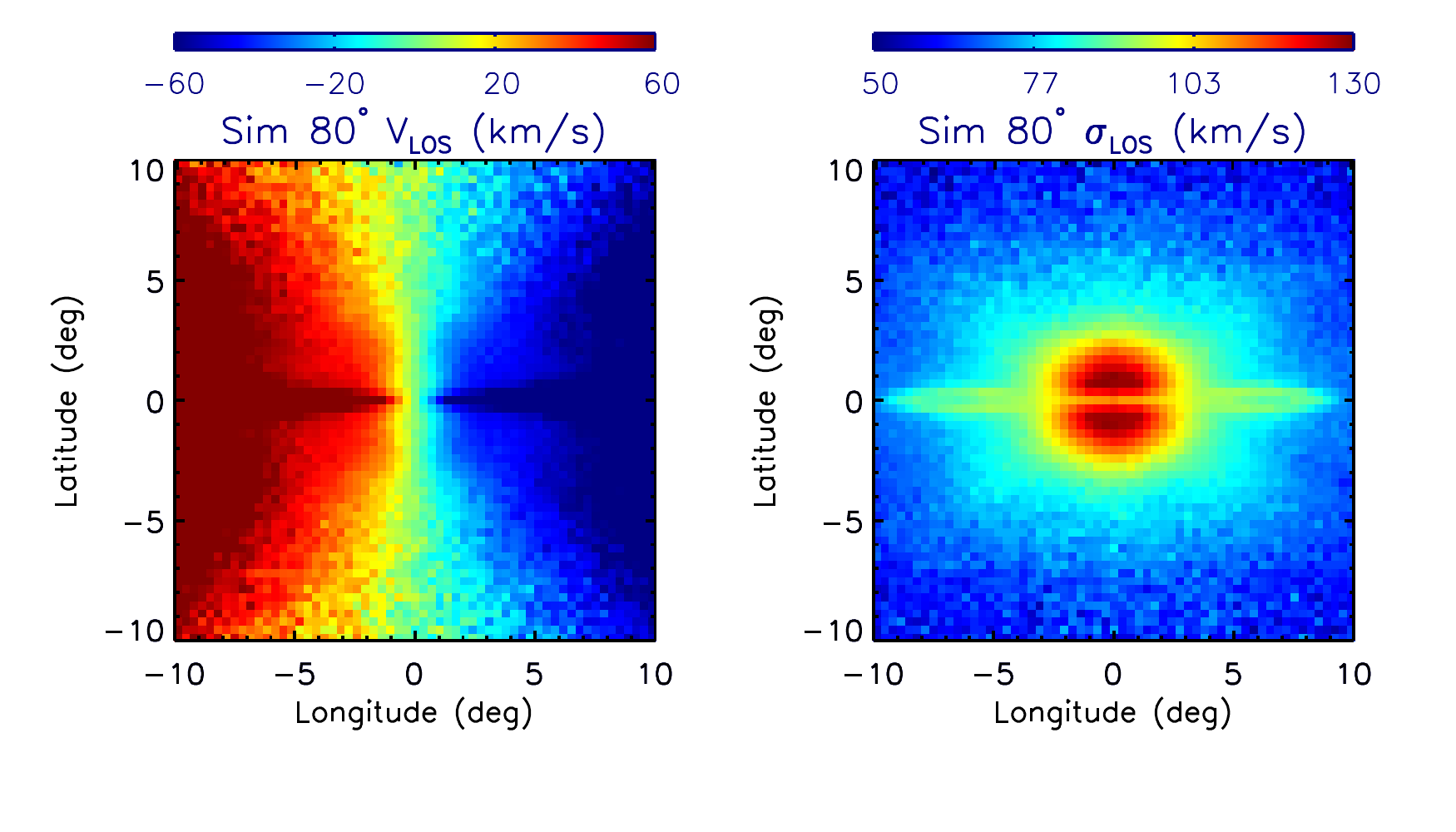}\\  \includegraphics[scale=0.53,angle=0,trim=0.5cm 0.5cm 0.5cm 0cm, clip=true]{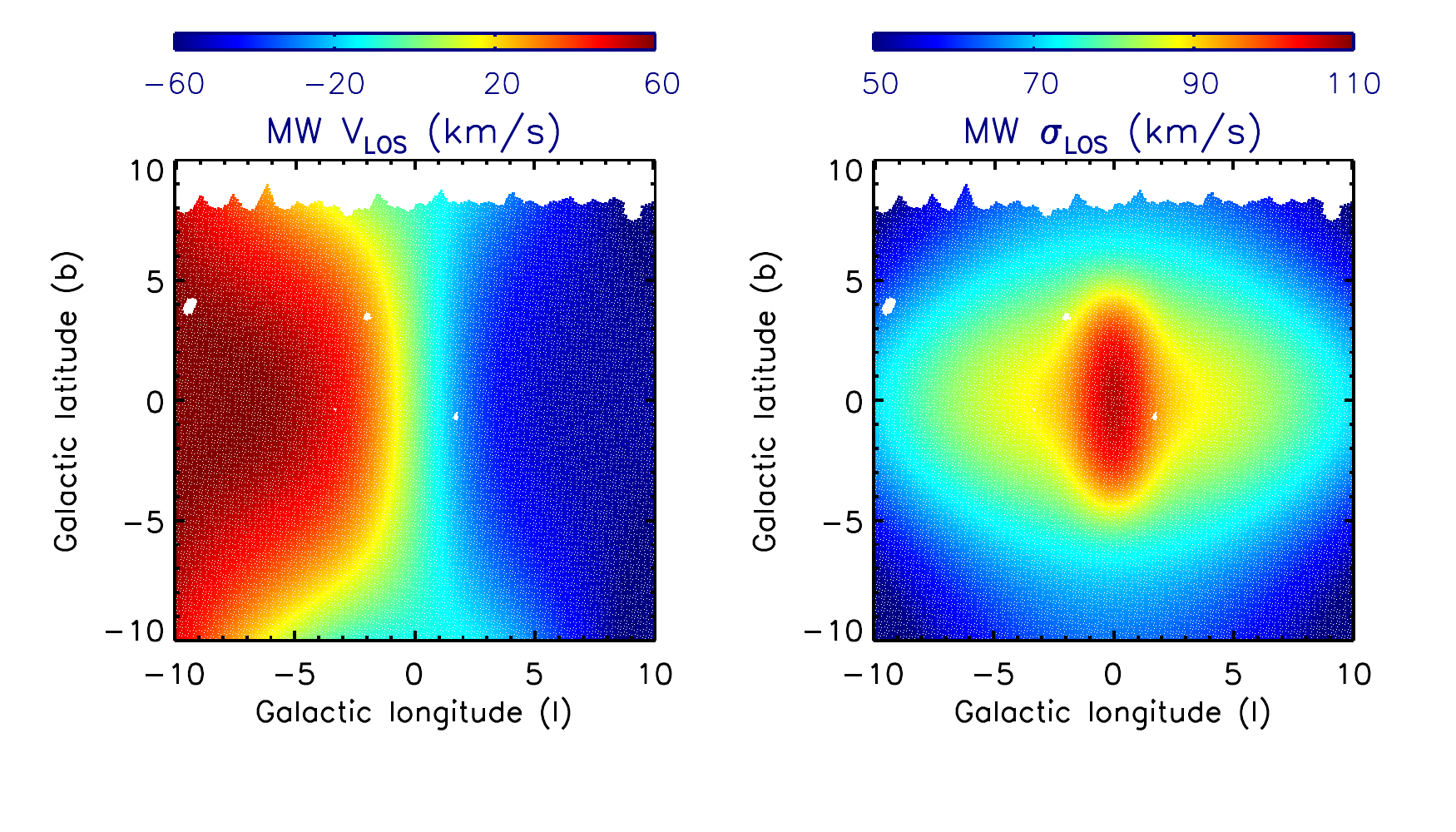}
\includegraphics[scale=0.53,angle=0,trim=0.5cm 0.5cm 0.5cm 0cm, clip=true]{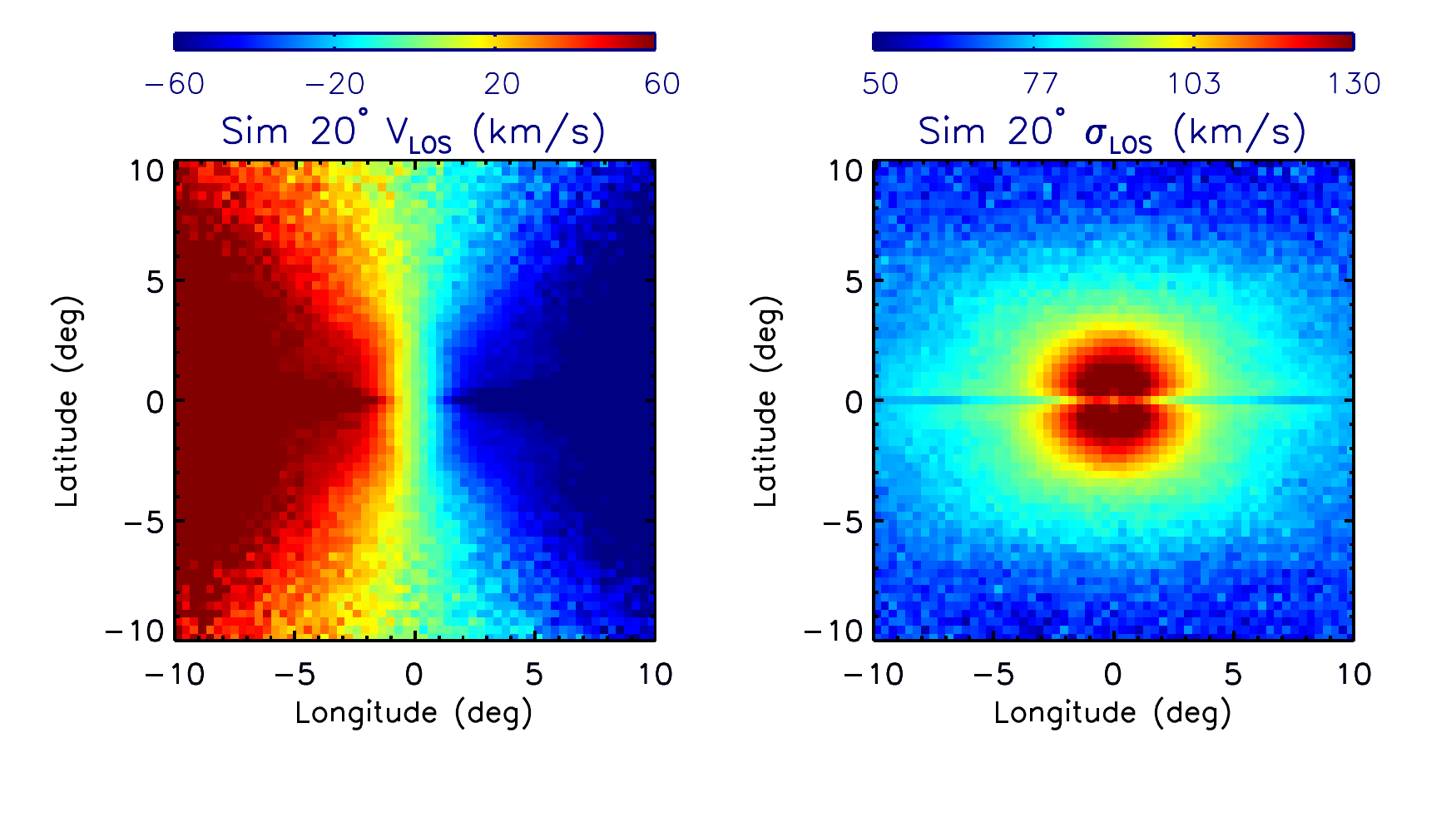}
}
\caption{The two upper left panels show the LOS velocity and velocity dispersion maps of NGC~4710 projected and oriented as the Milky Way bulge is seen from the Sun. The two lower left panels show the corresponding LOS velocity and velocity dispersion maps for the Milky Way bulge constructed evaluating each spatial resolution element of our NGC~4710 maps in Eq. 1 and 2 in \citet{zoccali+14}. The right side panels show the corresponding velocity maps for the simulation of \citet{cole+14} for bar orientation angles of 80$^{\circ}$ (upper panels) and 20$^{\circ}$ (lower panels).}
\label{fig:rvmap_mwlike_gibs}
\end{figure*}

The advent of spectroscopic and photometric surveys covering a large area of the Milky Way bulge are revealing properties that were previously inaccessible to us. One of the most recent observational findings has been that the Galactic bulge hosts a B/P or X-shaped structure. Such components are often found in external galaxies with a suitable orientation. Usually, a single deep image might already provide an answer to whether its bulge hosts this kind of structure. However, characterising the Milky Way bulge structure requires an entirely different approach based on the morphological reconstruction of an enormous amount of data where the distance estimation is obtained from the magnitudes of red clump giant stars \citep{zoccali+10, mcwilliam-fulbright-rich+10, mcwilliam-zoccali+10, nataf+10, saito+11, wegg-gerhard+13}. Such a detailed star by star reconstruction of the properties of the bulge are only possible in the Milky Way. Understanding bulges in external galaxies requires the interpretation of properties obtained from different techniques, such as image decomposition and spatially resolved kinematics, to be combined. Thus, providing a link between techniques used in external galaxies and in the Milky Way is of great importance.

\citet{zoccali+14} derived radial velocities for a sample of 6390 bulge red-clump stars from the GIRAFFE Inner Bulge Survey (GIBS). They constructed rotation curves at four different latitudes ($b=-2^{\circ}$, $-4^{\circ}$, $-6^{\circ}$, and $-8^{\circ}$) and interpolated between their fields to obtain the first rotation and velocity dispersion maps of the Milky Way bulge. We can use our maps of NGC~4710, where the entire kinematics could be measured directly, to compare to those of \citet{zoccali+14}. We thus transformed the radial velocity and velocity dispersion maps of NGC~4710 into a reference frame that is comparable to that of the Galactic bulge as seen from the Sun. We first transformed the photometric centre of NGC~4710 \citep[$\alpha = 12h49m38.8s$, $\delta=+15^{\circ}09'56.9''$; ][]{gadotti+12} to pixels using the astrometrical solution of our MUSE reconstructed image. The centre of NGC~4710 is found in our image at pixel positions x=141, y=189. We used these pixel values as the new reference centre and further applied a rotation of 64$^{\circ}$ in order to obtain a new reference frame (x', y') centred on NGC~4710 where the position angle of the major axis of the galaxy is zero. On the other hand, the pixel scale of MUSE of 0.2 ''/pix indicates that each pixel in our map would be equivalent to 0.016 \mbox{kpc} at the distance of NGC~4710 (16.9 \mbox{Mpc}). In the Milky Way, the bulge is located at approximately 8 \mbox{kpc} from the Sun, such that an angular size of 1 deg corresponds to 0.139 \mbox{kpc}. Thus, if the bulge of NGC~4710 would be located at 8 \mbox{kpc} from the Sun, each resolving element of our maps would have 0.016 / 0.139 = 0.11 $\mathrm{deg}/\mathrm{pix}$.  Applying this new scale we constructed new maps for NGC~4710 in a reference frame that can be compared directly to the Galactic coordinate system of the Milky Way.  

We have also included in this comparison the corresponding kinematic maps for the B/P bulge of the simulation of \citet{cole+14} (see Fig.~\ref{fov} and section 3). We applied a scaling factor of 1.2 to the model coordinates to scale its bar size of $\rm R_{bar}$=2.9 kpc to the bar size of the Milky Way\footnote{We note that the length of the Milky Way bar has been measured to be 4.4 kpc in the Galactic plane. However, since here we are scaling the inner parts of the bar that form the B/P bulge, we use the length of 3.5 kpc measured at larger distances from the Galactic plane \citep{wegg+15}.}, as described in \citet{ness+14}. For comparison purposes, the corresponding radial velocities and velocity dispersions have also been normalized so that the maximum radial velocity of all the bulges is comparable to that of NGC~4710 by applying a scaling factor of 0.48 to the velocities of the model and of 0.77 to the velocities measured for the Milky Way by \citet{zoccali+14}. Such spatial and velocity scaling transformations changes the reference systems but have no effect on the rotational patterns we are aiming to investigate here.

\begin{figure*}[]
\centering
\resizebox{\hsize}{!}{
\includegraphics[width=3.5cm,angle=0,trim=0.5cm 0cm 0.5cm 0cm, clip=true]{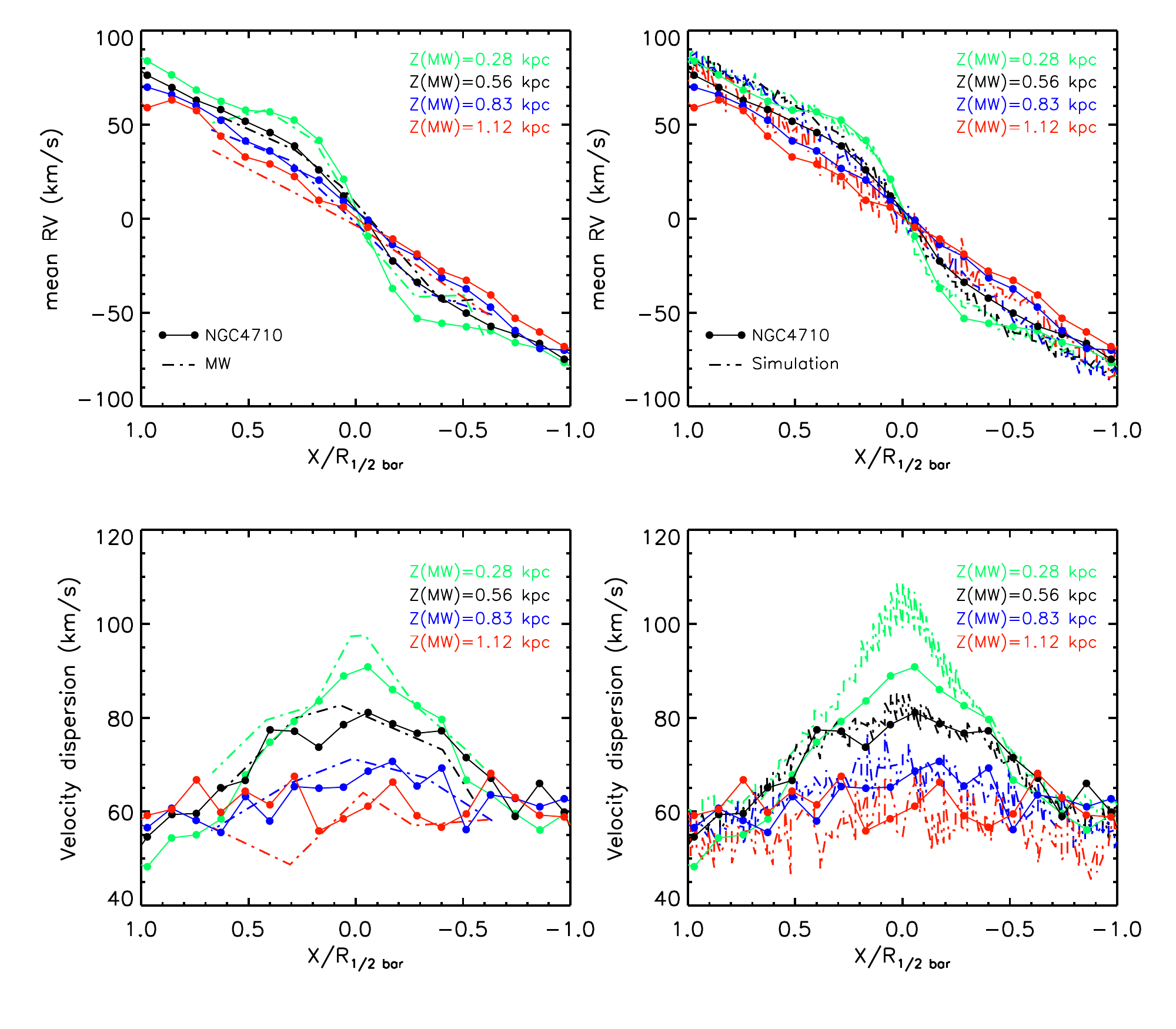}
}
\caption{The filled circles joined by solid lines show the mean LOS velocity and velocity dispersion curves of NGC~4710 at four heights from the plane Z = 0.28, 0.56 , 0.83, and 1.12 kpc. Dashed lines show the corresponding rotation curves of the Milky Way bulge from \citet{zoccali+14} (left panels) and of a disc galaxy simulation from \citet{cole+14} (right panels). Spatial coordinates of NGC~4710 and of the simulation have been scaled to match the Milky Way bulge bar scale length ($\rm R_{bar}$=3.5 kpc). The mean velocities and velocity dispersion of NGC~4710 and the simulation have been obtained in bins of 0.2 kpc, at the same scale heights of the Milky Way bulge measurements using the spatial scaling factor of 0.88 and 1.2 for NGC~4710 and the simulation, respectively. Velocities of the Milky Way bulge and of the simulation have been scaled to match the maximum velocity of the bulge of NGC~4710.}
\label{fig:rotcurves_mwlike_gibs}
\end{figure*}

This simple exercise allows us to visually compare the kinematical patterns of NGC~4710 with those of the Milky Way by the different Milky Way surveys. The position angle of the bar of NGC~4710 with respect to our LOS must be much closer to a side-on projection than for the case of the Milky Way, as suggested by the pronounced X-shape observable directly. This gives us the chance to investigate the effects of the bar orientation angle in the kinematic maps of the Milky Way bulge from \citet{zoccali+14}.

We can immediately see a similarity between the LOS velocity map of the bulge of NGC~4710 and that of the Milky Way in Fig~\ref{fig:rvmap_mwlike_gibs}. This is particularly true when considering that the apparent higher smoothness of the velocity dispersion map for the MW, presented in Zoccali+14, with respect to that NGC~4710 is a natural consequence of the different adopted techniques. For the Milky Way case, the map was obtained interpolating between the GIBS fields grid. In addition, it can be seen that the central peak in velocity dispersion that is found in the central region of the Milky Way bulge map is not seen with the same concentration and vertical elongation in the maps of NGC~4710. As already pointed out by Zoccali+14, while the presence of the central sigma peak is strongly supported by accurate velocity measurements of a large and statistically robust sample  (~450 individual stars), the extension of the peak is however poorly constrained due to the presence of only two fields in that region. Similar changes in the central velocity dispersion profiles are among the features identified in the LOS kinematics maps of \citet{iannuzzi15} and they are attributed to the effect of bar position angle. We see a similar change in the velocity dispersion maps constructed using different bar orientation angles in the simulation. Fig~\ref{fig:rvmap_mwlike_gibs} shows how the $\sigma$ map of the simulated B/P bulge becomes less vertically elongated when seen at (or close to) side-on projection, which is the case of NGC~4710. However, in this case the elongated central feature is not as pronounced as seen in the velocity dispersion map of the Milky Way bulge. 

To compare the velocity dispersion of the innermost fields of the Milky Way and NGC~4710 in a more quantitative way we now look at their 1-D LOS velocity and velocity dispersion profiles. In particular, we investigate their variations as a function of height. In order to do this comparison, we have scaled the (x', y') spatial coordinates of NGC~4710 by a factor 0.88 corresponding to the ratio between the bar sizes of the Milky Way $\rm R_{bar}$=3.5 kpc and NGC~4710 of $\rm R_{bar}$=3.9 kpc. With these transformations we have all three datasets spatially scaled to the Bulge of the Milky Way. The velocity profiles for NGC~4710 are then constructed at the four Milky Way heights presented in \citet{zoccali+14}, $\rm Z_{MW}=$ 0.28, 0.56, 0.83, 1.12 kpc, which they used to interpolate the velocity maps. The 1-D rotation and velocity dispersion profiles shown in Fig.~\ref{fig:rotcurves_mwlike_gibs} for the corresponding heights in NGC~4710 have a remarkable similarity to those of the Milky Way bulge and the differences seen between the inner regions of the $\sigma$ maps of the Milky Way and NGC~4710 are not as evident in the 1-D curves.  This suggests that the vertical elongation observed in central region of the $\sigma$ map of the Milky Way bulge might not be real, as suggested by \citet{zoccali+14}, but instead it is an artefact from the plane interpolation method when including the $\sigma$-peak observed at $b=-2$. In this case, the $\sigma$-peak in the Milky Way bulge would be limited to Galactic latitudes $|b|<2$ ($\sim$0.28 kpc). This is also in good agreement with the conclusions of \citet{valenti+15}, where the high stellar density peak of the central Milky Way bulge closely follows a more axisymmetric $\sigma$-peak than the one presented in the GIBS velocity dispersion maps.

It is clearly seen in Fig.~\ref{fig:rotcurves_mwlike_gibs} that the radial velocity profiles become steeper towards the galactic plane. This same effect was seen in the Milky Way bulge using the BRAVA survey \citep{howard+09} and further confirmed with the inclusion of the rotation profile at $b=-2^{\circ}$ by the GIBS survey \citep{zoccali+14}. The simulated B/P bulge shows the same behaviour as in the Milky Way and NGC~4710. \citet{combes+90} and \citet{Athanassoula+02} among others have suggested that such a rotation pattern would be expected for a B/P bulge when the bar has a non-zero position angle with respect to the Sun-Galactic centre line-of-sight and should not be immediately interpreted as evidence for the need of an additional component to reproduce the observed rotation curves. Indeed, the light decomposition of NGC~4710 itself and the very minor deviation from cylindrical rotation found in the previous section show no evidence for an additional bulge component besides the B/P bulge. Furthermore, the pure-disc N-body models of bars that have been used to evaluate the rotation curves of the bulge show the same behaviour without the need to include a classical bulge component \citep{shen+10,zoccali+14}. Thus, the fact that the rotation curve of the bulge of NGC~4710 and the Milky Way are so similar suggests that both of them are dominated by a similar B/P bulge structure. This is strongly supported by the detection of a vertical metallicity gradient along the B/P bulge of NGC~4710 that is comparable to the one measured in the Milky Way bulge. 

The existence of a spheroidal bulge component in the Milky Way in addition to the B/P bulge, has been suggested based on the distance distribution of RR Lyrae \citep{dekany+13}, the bi-modality in the metallicity distribution \citep{hill+11}, and the different kinematics seen in metal-poor and metal-rich bulge stars \citep{babusiaux+10}. Furthermore, \citet{ness+14}, \cite{vasquez+13}, and \cite{rojas-arriagada+14} found evidence for the X-shape bulge of the Milky Way to be only traced by the metal-rich bulge stars. The properties of the bulge of the Milky Way are in general investigated by selecting red-clump stars from the colour-magnitude diagram at a given line-of-sight. Metallicity and kinematics are then obtained individually from the spectra of each red-clump star and used to construct the total metallicity distribution. If the metal-poor component, that shows different kinematics (and spatial distribution), is much less significant in number compared to the dominant stars in the B/P, then the integrated-light properties could be dominated by the B/P component and would not be detectable in the integrated spectrum. This highlights the importance of understanding the detailed properties of the Milky Way bulge, where kinematics and chemical abundances can be investigated on a star-by-star basis, perhaps hidden in the integrated light of external galaxies.  The origin of the apparent different properties between metal-poor and metal-rich stars in the Milky Way bulge, i.e. the presence of a classical bulge, remains to be fully understood. From our results we see that the integrated light of the bulge of NGC~4710, which is dominated by a B/P component, is consistent with the dominant structure, kinematics, vertical metallicity gradient, and stellar ages of the Milky Way bulge.


\section{Summary and Conclusions}
In this study we have investigated the properties of the bulge of the edge-on galaxy NGC~4710 using MUSE spectra. The bulge of this galaxy shows a strong X-shape profile. A light-decomposition study provides no evidence for the presence of a classical bulge in the inner regions of NGC~4710, showing instead that the integrated light from this galaxy is dominated by a bar and the associated B/P bulge out of the plane of the disc. 

We have constructed the LOS velocity and velocity dispersion maps of the bulge, finding that the bulge rotates cylindrically, with a $\delta^{n}_{CL}=0.9$, based on the definition from \citep{Saha+13} where a value of $\delta^{n}_{CL}=1$ corresponding to a bulge with perfect cylindrical rotation and values of $\delta^{n}_{CL}<0.75$ define bulges with non-cylindrical rotation. The slight deviation from perfect cylindrical rotation is most likely due to the position angle of the bar being close to - but not perfectly - side on.
The velocity dispersion map shows a central peak in velocity dispersion which appears asymmetric with respect to the major axis of the galaxy, thus suggesting a non-perfectly edge-on view of NGC~4710.

We investigated the stellar population properties of the B/P bulge of NGC~4710 by performing full spectral-fitting in 4 fields along its the minor axis and find that the dominant population of the B/P bulge is $\sim$9 Gyr and shows no variation as a function of distance from the plane. The mean metallicity of the bulge decreases consistently at increasing heights from the plane. We measure a vertical metallicity gradient of 0.35 dex/kpc up to 1.5 kpc from the plane of the galaxy.

We scale the maps of NGC~4710 to a system that can be compared directly the Milky Way bulge and compare the resulting maps with those constructed by \citet{zoccali+14} based on interpolation of the GIBS survey rotational curves. In this comparison we found that the rotation map of NGC~4710 and the one of the Milky Way show a remarkable similarity. Although the velocity dispersion map of NGC~4710 appears noisier than that of the Milky Way bulge due to the way the maps are constructed, we see that the velocity dispersion map of the Milky Way bulge is more vertically elongated in the center than the one seen in NGC~4710. We see a similar change in the velocity dispersion maps constructed using different bar orientation angles in the simulation from \citet{cole+14}. This is also in good agreement with the features identified in the LOS kinematic maps of B/P bulges modelled by \citet{iannuzzi15}. We thus suggest that central increase in the velocity dispersion profile of the Milky Way bulge can be partially the consequence of the Milky Way bar and B/P bulge viewing angle of $27^{\circ}$ with respect to the Sun-Galactic centre line-of-sight. However, a comparison of the rotation and velocity dispersion curves at different heights from the plane, used by \citet{zoccali+14} to obtain the rotation maps, are very similar to those of the bulge of NGC~4710. These 1D rotation profiles are also in good agreement with the simulation of a galaxy hosting a pure B/P bulge. The differences identified between the inner regions of the $\sigma$ maps of the Milky Way and NGC~4710 are not seen in the 1-D curves. This suggests that the vertical elongation observed in central region of the $\sigma$ map of the Milky Way bulge could be an artefact from the plane interpolation method to account for the $\sigma$-peak observed at $b=-2$. Instead, the $\sigma$-peak in the Milky Way bulge would be limited to Galactic latitudes $|b|<2$ ($\sim$0.28 kpc).

On the other hand, the vertical metallicity gradient of 0.35 dex/kpc measured in the B/P bulge of NGC~4710 compares well to the one measured in the Milky Way bulge by photometric \citep{gonzalez+13} and spectroscopic \citep{zoccali+08, rojas-arriagada+14, ness-abu+13} surveys of $\sim$0.40 dex/kpc.

In the Milky Way, there is significant evidence showing that the oldest bulge stars do not follow the bar and B/P bulge spatial distribution. On the other hand, no spheroidal bulge component is found when performing an integrated light decomposition analysis of NGC~4710, thus being classified as a pure B/P bulge. If present in NGC~4710, a spheroidal bulge component would have to be much less dominant than the B/P to remain hidden from the light decomposition analysis. Despite this, our results show that the global chemodynamical properties of the bulge of NGC~4710, based on its integrated spectrum, are in excellent agreement with those of the resolved stellar populations of the Milky Way bulge, thus suggesting that these properties, i.e. the observed kinematic profiles and vertical metallicity gradient, are intrinsic to their B/P bulge component. Certainly, the ongoing and planned surveys based on multi-object spectroscopic facilities with high multiplex capabilities (APOGEE, 4MOST, and MOONS) where a large number of metal-poor stars can be mapped, might hold the key to understand the nature of the oldest, most metal-poor stars of the Galactic bulge. 

This study compares the global kinematics and morphological properties of the MW bulge to an external galaxy. This work is demonstrative of the potential to understand our own galaxy in context, via comparisons of its detailed stellar populations and kinematics from the rapidly growing large coverage spectroscopic and photometric datasets, with those of other galaxies observed with IFU instruments, such as MUSE.


\begin{acknowledgements}
We are grateful for the useful comments received from an anonymous referee. We warmly thank the ESO Paranal Observatory staff for performing the observations for this programme. MZ and DM acknowledge funding from the BASAL CATA through grant PFB-06, and the Chilean Ministry of Economy through ICM grant to the Millennium Institute of Astrophysics. MZ acknowledge support by Proyecto Fondecyt Regular 1150345. Support for this project is provided by CONICYT's PCI program through grant DPI20140066. DM acknowledge support by FONDECYT No. 1130196. VPD is supported by STFC Consolidated grant \# ST/J001341/1. The simulation used in this study was run at the High Performance Computer Facility of the University of Central Lancashire.
\end{acknowledgements}

\bibliographystyle{aa}
\bibliography{mybiblio_rev_full}

\end{document}